# Multimodal assessment of best possible self as a self-regulatory activity for the classroom

Batuhan Sayis
*Department of Information and Communication Technologies*
*Universitat Pompeu Fabra*
Barcelona, Spain
batuhan.sayis@upf.edu

Marc Beardsley
*Department of Information and Communication Technologies*
*Universitat Pompeu Fabra*
Barcelona, Spain
marc.beardsley@upf.edu

Marta Portero-Tresserra
*Department of Psychobiology and Methodology of Health Sciences*
*Universitat Autònoma de Barcelona*
Barcelona, Spain
marta.portero@uab.cat

***Abstract*—Best possible self (BPS) is a positive psychological intervention shown to enhance well-being which involves writing a description of an ideal future scenario. This paper presents a comparison of psychophysiological effects of a BPS activity that has been adapted for classroom settings and a time-matched control activity (NA). Thirty-three undergraduate students participated in the study that assessed state anxiety (State-Trait Anxiety Inventory, STAI), affect (Affective Slider, AS), and cardiac vagal activity (heart-rate variability, HRV) as an indicator of self-regulatory resource usage, at three time periods (PRE, DURING, POST). Results show that BPS led to a significantly greater increase in positive valence (DURING) and overall higher levels of cardiac vagal activity (HRV) compared to NA. These findings suggest that BPS has promising characteristics as a self-regulatory technique aimed at fostering positive affect and positively impacting self-regulatory resources. As BPS does not require expert knowledge nor specialized technology to administer, it may be a suitable activity for educators to use when teaching and having students practice self-regulation. This study presents evidence collected in a replicable multimodal approach of the self-regulatory effects of a brief BPS activity on undergraduate students.**

## I. Introduction

High levels of stress combined with underdeveloped coping mechanisms are contributing to a decline in university students' mental health [1,2]. A study involving undergraduate students found that self-regulation capacity predicted variance in levels of stress, mental health functioning, and psychological well-being [3]. The authors argue that universities should provide ongoing self-regulation training to students through formal courses. Likewise, a lack of self-regulation skills to manage stress, such as being able to identify one's own emotional states and modifying these states when necessary for a better adaptation to the environment, affects cognitive processes, well-being, and contributes to a decrease in academic performance [4]. A scoping review of academic literature on university student stress and mental well-being, found that skills-oriented interventions with supervised practice were the most effective, with mindfulness interventions and cognitive behavioral interventions producing the greatest change in mental health outcomes [5]. Furthermore, in their analysis of a survey study of university students' mental well-being, the authors write that students believe that "university teachers and their teaching practice have the potential to enhance and support student mental well-being" [2]. However, many educators lack knowledge and adapted resources to adequately teach and guide students through self-regulation practices [6,7].

Internet-based digital technologies for teaching and learning can play a role in supporting educator use of self-regulation activities in the classroom. For example, an internet-based digital technology for teaching and learning (ClassMood App) [8] contains an openly accessible database of self-regulation activities for classroom use. The database includes both activating and calming activities such as breathwork and mindfulness exercises. Empirically validated activities are highlighted with references included in activity descriptions. These activities can act as forms of supervised practice of skills-oriented self-regulatory interventions. Yet, empirical evidence is lacking for versions of these activities that have been adapted to be suitable for the classroom. A recent study involving children found that a 1-min deep breathing intervention reduced physiological arousal [11]. The ultra-brief intervention design served to increase reuse in school and home settings. Hence, there is empirical evidence to support self-regulation strategies such as cognitive reappraisal [9], progressive muscle relaxation and diaphragmatic breathing [10]. However, unlike the 1-min deep breathing intervention, evidence is lacking for variations of the former that have been adapted for educational settings.

Self-regulation interventions aim to address somatic and cognitive experiences. Interventions targeting the latter often strive to help one identify and alter negative beliefs, which can underlie maladaptive emotional responses [12]. A self-regulation intervention, based on positive psychology [13], that shows promise for classroom use is called best possible self (BPS). BPS involves having participants write a description of an ideal future scenario where they have reached desired goals, after working hard toward them [14,15]. It is a mental imagery exercise that includes positive reappraisal, which can promote the modification of negative self-beliefs, improve positive affect, self-awareness, optimism [16] and positive future expectations [17]. BPS does not require expert knowledge nor specialized technology to administer, and it has been found to be effective when performed both digitally (online) and in-person (handwriting) [15]. Variations of BPS have also recently been administered by robotic well-being coaches [18,19]. Further, BPS has been

This work has been partially funded by Universitat Pompeu Fabra Initiatives (Planetary Wellbeing: PLAWB00321, PLAWB00322; PlaCLIK: E2022014338), Erasmus+ program (2021-1-ES01-KA220-SCH-000032801, RemixED), Spanish Ministry of Science and Innovation (PID2020-112584RB-C33 granted by MICIN/AEI/ 10.13039/ 501100011033) and European Union-NextGenerationEU, Ministry of Universities and Recovery, Transformation and Resilience Plan, through a call from Universitat Pompeu Fabra (Barcelona).

shown to enhance well-being across diverse populations [20] with even a single session resulting in improved well-being, positive future expectancies, and positive affect [21]. Studies have found benefits to the use of BPS in primary schools [22], high schools [23], and universities [24]. Studies evaluating BPS typically use self-reports, questionnaires, or scales to assess well-being (positive and negative affect, life satisfaction), optimism, and symptoms of depression [25]. A recent study evaluated the psychophysiological effects of BPS by examining cardiovascular markers of challenge and threat (e.g., cardiac output; liters of blood pumped by the heart in 1 min) [26]. The study evaluated the impact of a 16-min BPS activity on university students before a stress-provoking task. Results showed that this intervention led to an increase in adaptive coping especially in subjects with trait anxiety. Research could not be found that evaluates the psychophysiological effects of a brief BPS activity adapted for university classroom use [27].

In the context of multimodal assessments of self-regulation activities (SRA), [28] created a study protocol for comparing the efficacy of two self-regulatory breathing practices. The protocol manages carryover effects so that two 5-min practices could be assessed in a single session collecting multimodal data. The protocol involving a physiological measure (heart rate variability, HRV) and psychological measures (anxiety, arousal, and valence), involves rest, reactivity, and recovery periods. These periods follow the 3Rs of HRV [29,30] with rest denoting the baseline; reactivity the change between baseline and a specific event; and recovery the change between the event and a post-event measure. The study authors based their hypothesis on detecting changes in cardiac vagal activity (CVA), also referred to as vagally mediated HRV. CVA is thought to be an indicator of how efficiently self-regulatory resources are mobilized and used [31,29]. Furthermore, the authors identified the HRV parameter, RMSSD or root mean square of the successive differences, as an index of CVA. The authors argued that RMSSD has been shown to return to baseline immediately after the performance of some self-regulatory exercises suggesting that risk of carryover in within-subject designs is low with this parameter [28,32]. An objective of our work is to compare multiple self-regulatory activities for the classroom; therefore, we have conceptually replicated the protocol of [28].

A modification made to the protocol includes the use of a Vanilla Baseline (VB). VB is an alternative to resting forced relaxation, where subjects perform a task requiring sustained attention but with minimal cognitive load [33]. This type of activity is similar to tasks used to evaluate attention and/or visual working memory [34] and better reflects a students' classroom experience. Additionally, following the approach of [35], we designed the presentation of activities to participants in a manner that mimics how an educator would use such materials (i.e., presenting a video or digital instructions that have been adapted for classroom use).

All in all, this study contributes a multimodal assessment of the effects of a brief BPS activity (BPS) and control activity (NA) on student anxiety, affect (arousal and valence), and self-regulation (cardiac vagal activity) in a manner that enables a comparison with other SRA adapted for classroom use. An aim of this work is to contribute to efforts that provide students with opportunities in the classroom to learn about and practice self-regulation techniques. This work involves exploring how to improve the presentation and practice of techniques so that educators can identify which techniques may be appropriate for their classroom contexts (i.e., accounting for student characteristics and the challenging educational situations students face). In this study, we hypothesize that the BPS condition compared to the NA condition will (1) induce more positive affect (valence and arousal) as reported in previous studies on BPS [17]; (2) not differ in anxiety ratings as neither condition is expected to stress participants nor are stressor tasks being used in the study design from which participants must recover from; (3) lead to an increase in CVA indicating a greater self-regulatory effect [29] and thereby provide evidence for the use of BPS as a SRA for the classroom.

## II. Method

### A. Participants

Sample size calculations were informed by similar research where a self-regulatory session was compared with a control condition using a within-subject design [28,32]. Based on an assumption of medium effect sizes of micro-interventions on psychological indicators of health [26], as well as medium sized effects in education research [36], a G*Power a priori power calculation for a 2x3 repeated-measures ANOVA (within subject factors (NA, BPS); within subject factor time (PRE, DURING, POST)) to detect a medium effect size $f = 0.3$, power $(1-\beta) = 0.80$, correlation among repeated measures = 0.50, provided an estimated sample size of 31. Participants were recruited via flyers at the university campus and screened during the application process for factors that could affect the results (e.g., chronic heart issues or respiratory conditions). All 40 participants enrolled in the study, gave written informed consent before participation, and received 15 euros for their participation. Data from 7 participants were excluded from the analysis: 6 due to technical issues (electrode movement on signal integrity) during the HRV analysis and one due to a misuse of self-report measures. The final sample was 33 participants (18 female, 15 male) from 18 to 25 years old ($M = 19.54 \pm 1.71$) (20: 1st year students, 6: 2nd year students, 3: 3rd year students, 3: 4th year students).

### B. Materials and Measures

*1) Study Design:* All participants in the within-subject design joined 2 lab sessions held on separate days to increase the number of SRA that could be assessed and better control for carryover effects between activities [28]. Each session was approximately 1 hour and followed the same protocol as shown in Figure 1 in which two SRA (BPS or alternatives) and one control activity were performed. Each activity had a duration of 2 min. The duration of 2 min was selected as short activities present lower barriers to use by educators and activities as brief as 1-min have been found to have positive physiological effects [11]. The control activity (NA) was performed on day 1 and BPS on day 2. The assignment of the order of the activities within each session was carried out randomly to counterbalance the order effect of the activities.

*2) Writing about best possible self (BPS)*: BPS is often adapted for the context being studied. In studies with undergraduate students, themes differ broadly (e.g., social, health, academic, career), as do the length, intensity, magnitude, and format of the BPS activity [25]. Reference [24] used a BPS task that involved thinking (but not writing) of one's ideal academic future life and suggested that 2 min is sufficient for participants to imagine an ideal academic future life. For this study, we created a brief BPS activity (BPS) that included 1 min of thinking and 1 min of writing. A near future event was selected to reduce needed thinking time. Participants received the following instructions "Think about your life at the end of this course year. Imagine that everything has gone as well as possible. You have worked hard and succeeded at accomplishing all your goals. Think of this as the realization of all your goals this year. Now write about what you imagined." The task was completed while sitting down, using a pen and paper. During the familiarization phase of the session, participants were shown BPS instructions on a screen and prompted to ask questions if they did not understand.

*3) Neutral Activity (NA):* In a study of autonomic effects of a breathing exercise, [32] used a TV documentary ("Abenteuer Forschung" [Research Adventures]) about research discoveries related to space to serve as a control activity. The video was found to have emotionally neutral ratings. To match the 2-min duration of the self-regulatory activities in this study, different 2-min parts of this same documentary were shown on day 1 and day 2 and Spanish subtitles were added.

*4) State-Trait Anxiety Inventory (STAI)–Perceived Stress:* State-Trait Anxiety Inventory (STAI) consists of a self-report questionnaire to measure symptoms of anxiety and propensity to be anxious. A short form consisting of two subscales with 6-items each of the inventory was used in this study [37]; first the State Anxiety Scale (STAI-S) which evaluates the current state of anxiety; and second, the Trait Anxiety Scale (STAI-T) to evaluate stable aspects of anxiety proneness. Participants answered the STAI-T scale 15 min before the first SRA of each session. Participants answered STAI-S just before and after each of the experimental activities were performed in each session. A detailed description of the questionnaire, scoring and interpretation are described in [38].

*5) Affective slider–Perceived emotional arousal and valence:* Affective slider (AS) is a questionnaire to measure pleasure and arousal levels. It consists of continuous Likert scales with icon images on either ends to measure the levels of arousal ("How active do you feel?") and valence ("What is your mood?") [39]. All participants answered AS just before and after each of the experimental activities were performed in each session.

*6) Vanilla baseline:* Based on [33], participants were asked to observe a video screen and count the number of times a designated color occupies a 10x12 cm rectangle on the screen. The color of the rectangle changed every 6 s (1 s separation between each color), and the duration of the task was 2 min.

*7) Cardiac vagal activity indexed via heart rate variability*: HRV was measured via an electrocardiogram (ECG) signal acquired from a wearable (Superhero Collar) that has been validated to capture physiological data during the performance of full-body movements [40]. The wearable records acceleration via an accelerometer sensor, electrodermal activity (EDA) via electrodes placed on the shoulders, and ECG signal with electrodes on the chest. Data from the Superhero Collar was acquired in an "offline mode" to the internal memory of a Biosignal Plux device (PLUX Wireless Biosignals S.A.) at the sampling rate of 500 Hz. Open Signals software was used in a laptop to verify the correct placement of the electrodes prior to signal acquisition.

## C. *Procedure*

A university research ethics committee approved the study protocol (No. 226/2021). Participants were asked to not consume coffee or alcohol, smoke, or do high intensity physical activity prior to coming to the experimental sessions [30]. In the first session, participants were given a consent form and were verbally informed that signing the consent form would establish their consent to participate in the study. Participants were then introduced to the forthcoming sequence of activities (see Figure 1) that was to be repeated three times. Participants then sat in a chair facing a monitor and keyboard. The monitor presented the trial tasks and activities that were run on PsychoPy (v2021.2.3). Participants read a welcome message, completed a survey that collected demographic information and responses to questions related to exclusion criteria (medical conditions) and possible confounds, and completed a measure of trait anxiety (STAI-T). As the welcome survey was being completed, two neurotechnological devices for collecting physiological data were placed on participants. The first is the Superhero Collar which is described in Materials and Measures. The second device is a commercial, in-ear wearable Cosinuss One (Cosinuss Company, Munich). Data from the Cosinuss One is not being reported in this paper. To synchronize PsychoPy and data sources from these wearables, the event-based synchronization approach was followed with a universal clock

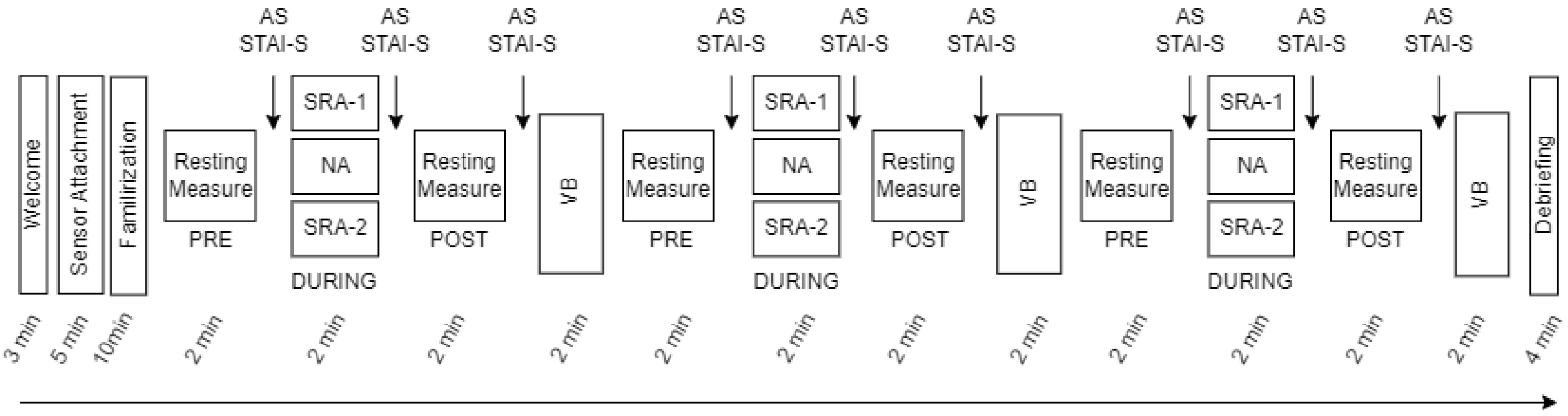

Fig. 1. Diagram of the experimental design

TABLE I. DESCRIPTIVE STATISTICS (NA VS BPS)

| | NA | | | BPS | | |
|---|---|---|---|---|---|---|
| | ***PRE*** | ***DURING*** | ***POST*** | ***PRE*** | ***DURING*** | ***POST*** |
| Arousal | (2.75 ± 0.83) | (2.94 ± 0.76) | (2.44 ± 0.90) | (2.75 ± 0.73) | (3.30 ± 0.70) | (2.91 ± 0.73) |
| Valence | (3.61 ± 0.48) | (3.49 ± 0.51) | (3.36 ± 0.50) | (3.40 ± 0.53) | (3.59 ± 0.52) | (3.43 ± 0.57) |
| lnSTAI-S | (3.36 ± 0.18) | (3.44 ± 0.21) | (3.33 ± 0.17) | (3.43 ± 0.26) | (3.47 ± 0.24) | (3.42 ± 0.23) |
| lnRMSSD | (3.80 ± 0.59) | (3.66 ± 0.60) | (3.78 ± 0.60) | (3.89 ± 0.54) | (3.84 ± 0.53) | (3.94 ± 0.53) |
| STAI-T | (43.58 ± 8.71) | | | | | |

Abbreviations: lnRMSSD, root mean square of the successive differences (log); lnSTAI-S, State-Trait Anxiety Inventory (state) (log); STAI-T, State-Trait Anxiety Inventory (trait) presented on a tablet PC that was placed in the experimental setting visible to the experimenter [41]. Next, participants were familiarized with the specific tasks and activities: self-report scales (STAI-S, AS), resting task, color counting task (VB), and SRA. For each, participants were first shown a screen with instructions, then shown the screen with the task as it would appear during their performance and prompted to ask questions if they did not understand the task. The resting task involved having participants look at a black screen with a fixed white cross in the center of the screen [42]. After the familiarization phase, the progression of the trial was controlled by the participant on questionnaire tasks, and advanced automatically based on pre-set times for the resting, recovery, and activity tasks. Upon finishing the final VB period, participants completed a questionnaire related to their views on the appropriateness of the adapted activities for use in a classroom setting. Wearables were then removed, participants debriefed and thanked for their participation.

### *D. Data Analysis*

Participants completed a total of 6 activities (self-regulatory and control) across 2 days. Physiological data collected included HRV, EDA, and body temperature. This paper presents an analysis of the control (NA) and BPS (BPS) activities and focuses on self-report measures (AS; STAI-S) and HRV (RMSSD) in line with the work by [28]. Statistical analyses were computed using SPSS (v29). HRV was computed from raw ECG signals acquired from the Superhero Collar. The signal data was imported into Kubios (version 2.2), and samples were analyzed manually for artifacts. All features were extracted in the time and frequency domains, and nonlinear indices that the spectrum of Kubios includes. Data were checked for normality and outliers. As RMSSD data were non-normally distributed, a log-transformation was applied (lnRMSSD), as is often recommended for HRV research [30]. As STAI-S data were also mostly non-normally distributed, similar to RMSSD, we applied a log-transformation (lnSTAI-S). For AS, 1 outlier was removed from valence ratings. AS valence and arousal data were mostly normally distributed. A series of repeated measures ANOVA with Greenhouse–Geisser correction were conducted, with conditions (BPS vs. NA) and time (PRE, DURING, POST; referring to resting, reactivity, and recovery) set as independent variables, with valence, arousal, perceived stress intensity (STAI-S) as self-report dependent variables, and lnRMSSD as an HRV-dependent variable indexing CVA. Adjustments for multiple comparisons were done with Bonferroni correction.

## III. RESULTS

Descriptive statistics can be found in Table 1, and in Figure 2 to Figure 5 (* represents $p \leq 0.05$, ** represents $p \leq 0.01$, *** represents $p \leq 0.001$).

Our first hypothesis was that BPS would induce more positive affect (valence and arousal) as reported in previous studies on BPS (Heekerens & Eid, 2021). Firstly, results show that no significant main effect of the condition on arousal levels was found, $F(1, 32) = 3.91$, $p = 0.056$. However, this result on the main effect was qualified by an interaction between condition and time, $F(1.74, 55.76) = 5.94$, $p = 0.002$. Checking the simple main effects reveals how arousal levels of BPS in DURING and in POST are different from (significantly higher than) arousal levels of NA in DURING $p = 0.032$ and in POST $p = 0.005$ (see Figure 2). In sum, BPS

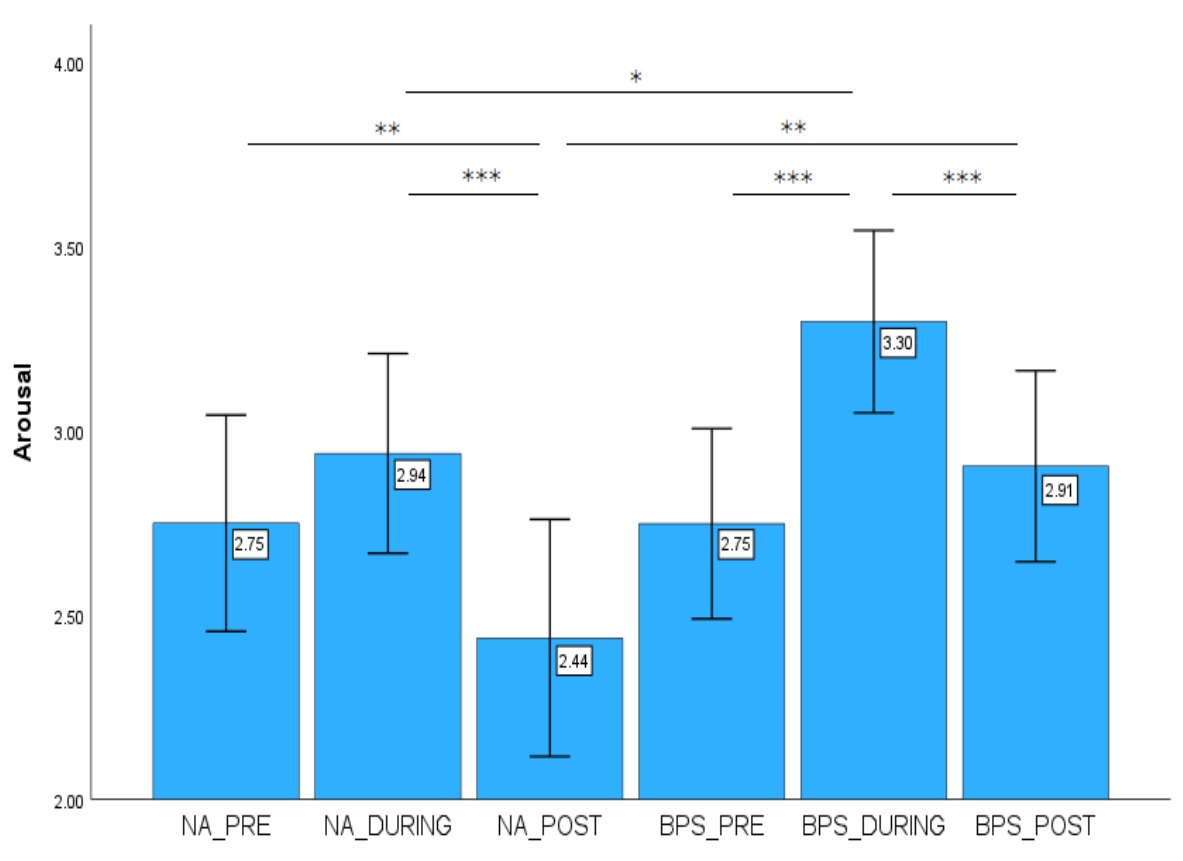

Fig. 2. Descriptive Statistics (Arousal)

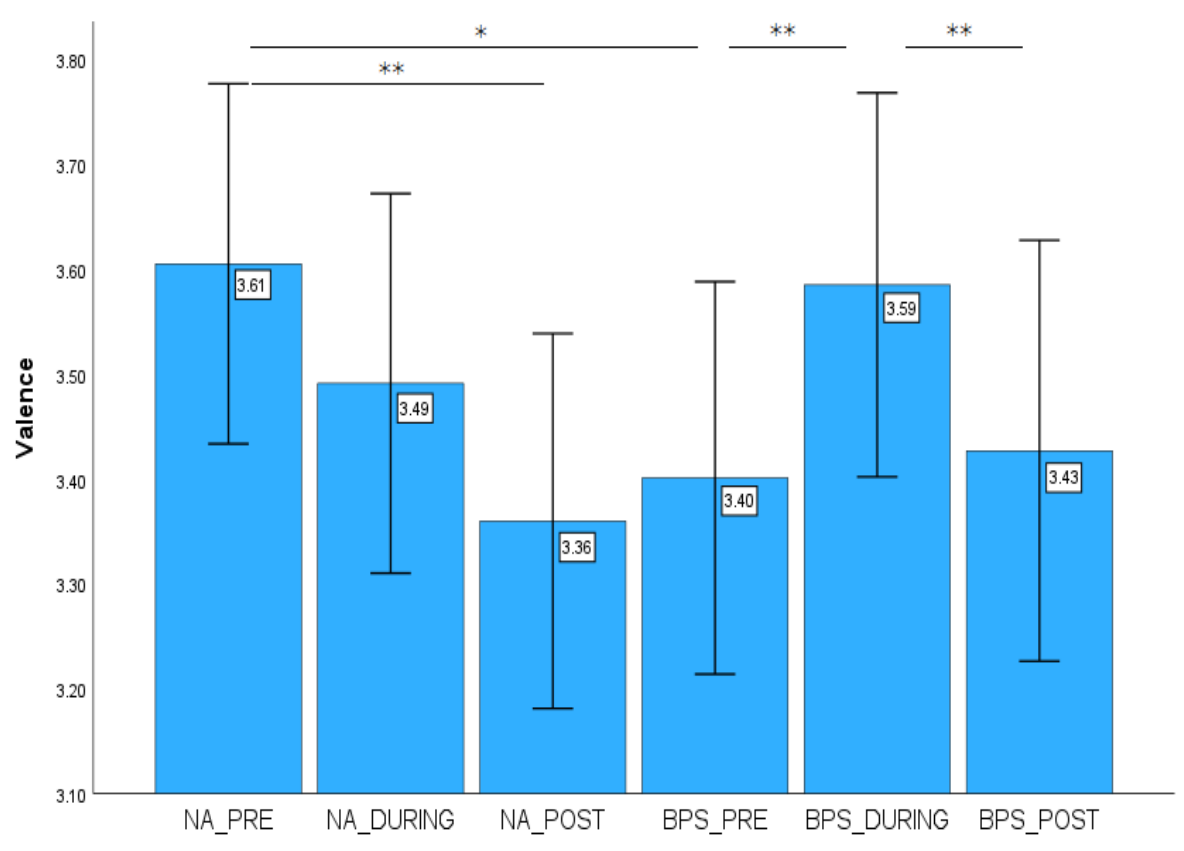

Fig. 3. Descriptive Statistics (Valence)

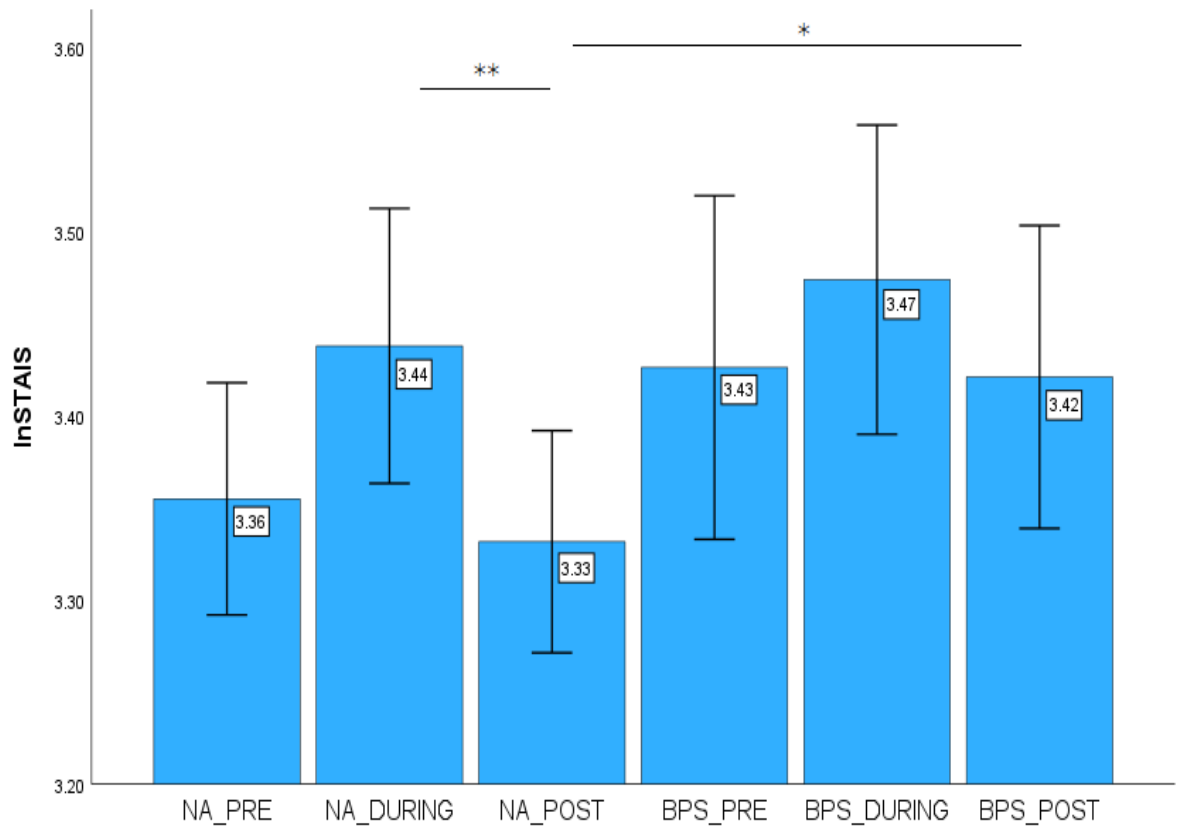

Fig. 4. Descriptive Statistics (lnSTAIS)

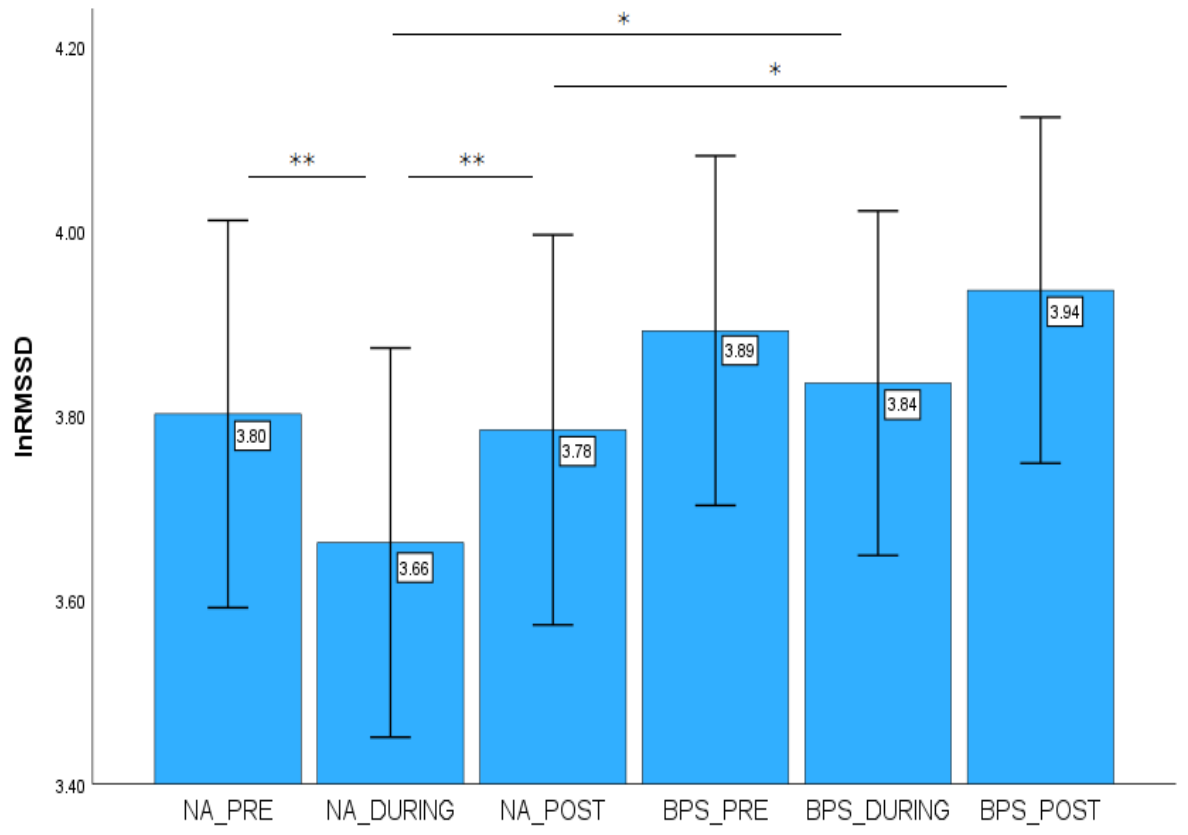

Fig. 5. Descriptive Statistics (lnRMSSD)

leads to significantly higher levels of arousal than NA once the activity has been initiated.

Secondly, results show that no significant main effect of the condition on valence levels was found, F(1, 32) = 0.38, p = 0.84. However, this result on the main effect was qualified by an interaction between condition and time, F(1.58, 50.66) = 7.47, p = 0.003. Checking the simple main effects reveals how valence levels of BPS in PRE are different from (significantly lower than) valence levels of NA in PRE p = .018 (see Figure 3). Simple main effects also show how valence levels of BPS in DURING and in POST are similar to (no significant difference) valence levels of NA in DURING, p = 0.29 and in POST p = 0.49. Given the lower starting level of valence in BPS, results suggest that BPS had a more positive effect on valence. In sum, BPS brought participants to a significantly higher valence state from PRE to DURING while NA worked in an opposite direction with significant negative effects on valence.

Our second hypothesis was that BPS would not differ in anxiety ratings as neither condition was expected to stress participants nor were stressor tasks being used in the study design from which participants must recover from. Results (lnSTAI-S) support this hypothesis as no significant main effect of the condition was found, F(1, 32) = 2.41, p = 0.13. Moreover, there was no interaction effect between time and condition F(1.60, 51.46) = 0.81, p = 0.42. In sum, BPS does not differ from NA in terms of perceived anxiety (see Figure 4). Our third hypothesis was that BPS would lead to an increase in CVA indicating a greater self-regulatory effect [29]. Results (lnRMSDD) support this hypothesis as a significant main effect of the condition was found, F(1, 32) = 4.52, p = 0.041. The overall InRMSSD level was found to be higher in BPS than in NA (see Figure 5). However, there was no interaction effect between time and condition, F(1.86, 58.82) = 1.30, p = 0.27. In sum, BPS, leads to higher levels of CVA indicating a greater self-regulatory effect than NA.

## IV. Discussion

The present study was designed to assess the effect of a brief BPS activity (BPS) on student anxiety, affect (arousal and valence), and self-regulation (cardiac vagal activity) in comparison to a control condition (NA). The gathered evidence is to contribute toward efforts exploring how to improve the presentation and practice of self-regulation techniques so that educators can identify which techniques may be appropriate for their classrooms. Results suggest that BPS in comparison to NA leads to significantly higher levels of arousal and valence once the activity has been initiated; does not lead to increases in perceived anxiety; leads to higher levels of CVA which indicates a self-regulatory effect; and, thus, can be considered for use by educators as an evidence-based self-regulatory activity for the classroom. In relation to effects on levels of arousal and valence, BPS brings participants to significantly higher active states in DURING and POST compared to NA. These findings are consistent with previous literature showing that both the parasympathetic nervous system (PNS) and sympathetic nervous system (SNS) increase during emotional tasks [43]. For example, a study using physiological measures found that writing, regardless of the topic, increases SNS activity among subjects [44]. Thus, it is not surprising to see a significant change in BPS in DURING and POST compared to NA. These findings are also consistent with lnRMSSD results as arousal plays a role in self-regulation [45]. Moreover, no increase in perceived anxiety when comparing BPS to NA, suggests that the increase in arousal seen in the BPS condition relates to participants experiencing more positive moods such as joy or excitement rather than anxiety as suggested by the circumplex model of affect [46]. This interpretation is consistent with the higher levels of valence and lnRMSSD results of the BPS condition as higher CVA is associated with more positive emotional experiences [29]. Accordingly, it should be noted that the decrease of CVA observed with BPS and its subsequent increase after completing the BPS task is in line with previous research, in which a cognitive task is thought to have such an effect [29]. However, a similar decrease and subsequent increase in lnRMSSD observed in NA is not in line with previous research [28] in which no changes in lnRMSSD could be observed in a similar control condition. A possible explanation for this could be the language of the video which differed from the native language of participants and the addition of subtitles to facilitate understanding of the video. These changes may have affected the emotional response to the video. Both NA_DURING and BPS_DURING lnRMSSD levels go back to PRE levels after each condition. While there is no significant difference between PRE_NA and PRE_BPS (where BPS is higher than NA) lnRMSSD levels, there is a significant difference between POST_NA and POST_BPS (where BPS is higher than NA). This significant difference suggests that after BPS, a higher self-regulatory effect in participants is found. These results are in line with previous research where HRV levels are higher after expressive writing tasks [44]. Although, the writing task, sample profile, and HRV features that show this higher effect differ from our study.

## V. LIMITATIONS

Although results are promising, some important limitations need to be considered. (1) An alternative control condition should be considered for future studies. NA condition results were unexpected as NA was selected to be an emotionally neutral control condition [32, 28] but results suggest it led to negative effects on valence. (2) To facilitate the assessment of multiple activities in a single session, the choice of self-report measurements was limited to brief measures. Comparing results from these measures with more frequently used measures (e.g., PANAS) in BPS studies should be done with caution. (3) Significant differences in PRE measures of valence between conditions should be further investigated to see if there were carryover effects (i.e., from the negative valence resulting from the NA condition). (4) Data from a few participants had to be excluded from the analysis because of the quality of the data which reduced the final sample size. (5) Most participants were 1st year university students which creates a possible bias limiting the generalizability of results. (6) Self-report measures are subject to biases such as honesty, introspective ability, and interpretation of questions. (7) Although the study design (instruments, protocol) can be adapted to work in a classroom setting and with longer activities, the present study was conducted in a laboratory setting, with brief 2-minute activities, and performed individually rather than in groups which decreased its external validity but facilitated the assessment of multiple activities.

## VI. FUTURE WORK

Future research should tackle the limitations mentioned with new studies. This includes expanding the activities being evaluated and testing these activities in group or class settings. Future studies should also explore adaptations for neurodiverse populations and the effect of gender and personality traits. From a data analysis point of view, as a next step, HRV measurements can be extended to include other HRV features (e.g., SDNN, HF, LF) and coupled with physiological measurements such as EDA and body temperature for a more comprehensive evaluation. In this regard, Machine Learning techniques can be used for the fusion of multimodal data. Further, to balance out the limitations of the quantitative data, data from the questionnaire completed by participants related to their perception on the appropriateness of the adapted activities should be analyzed with qualitative data analysis techniques.

## VII. CONCLUSIONS

This work contributes a multimodal assessment of the effects of a brief BPS activity and control activity on student anxiety, affect (arousal and valence), and self-regulation (cardiac vagal activity) in a manner that enables a comparison with other self-regulation activities adapted for classroom use. Limitations and suggestions for improving the protocol are shared. Key findings from the evaluation of activities extend findings of previous work related to BPS and provide preliminary evidence to support the use of BPS as a classroom self-regulation activity as BPS was found to have a self-regulatory effect on participants from a psychophysiological point of view. The gathered evidence can also be added to the open database of ClassMood App that aims to support educator use of self-regulation activities in the classroom.

## ETHICAL IMPACT STATEMENT

A university research ethics committee approved the study protocol (No. 226/2021). All our research is within the EU Responsible Research and Innovation policy (RRI) (https://ec.europa.eu/research/innovation-union). RRI is defined as “a higher-level responsibility that aims to shape, maintain, develop, coordinate and, align existing and novel research and innovation-related processes, actors, and responsibilities with a view to ensuring desirable and acceptable research outcomes.’’[47]. One of the core requirements of RRI is privacy which is defined by The European privacy directive (95/46/EC), subsequent national legislation and the current review of this entire system with a view of developing a unified European General Data Protection Regulation. In this regard, all data used from participants in our experimental trials has followed the protocols defined by applicable EU law and ethical standards through the resources provided by the university. Each participant's ID related information and other information types (psychophysiological data, behavioral data) is associated with a unique alphanumeric ID. To protect the participant’s privacy, the alphanumeric ID file which is associated with participants’ IDs are stored separately from the rest of the documentation in a university-controlled password-protected disk. Only the researchers of the project have access to them, under the supervision of the principal investigator. Regarding the generalizability of the research, we have also explored and discussed the results with psychologists on how this work can better meet the needs of disadvantaged students, and support mental literacy and self-regulation. Owing to the space limitation of this paper, the details of this discussion have been omitted.

## ACKNOWLEDGMENT

We would like to express our gratitude to Prof. Narcis Pares from Universitat Pompeu Fabra, and psychologists Maria Angeles Mairena Garcia de la Torre and Maria Diez Juan from Hospital Sant Joan de Déu for their contributions to the research. We would also like to thank all research participants who generously gave their time and effort to this project.